\documentstyle[prd,aps,preprint,tighten,epsfig]{revtex}

\begin{document}

\title{\bf{Corrections to Tribimaximal Mixing from Nondegenerate Phases}}

\author{Y.F.~Li$^{\,a,b\,}$\footnote{E-mail:
\texttt{lyfeng@mail.ustc.edu.cn}}  and
Q.Y.~Liu$^{\,a\,}$\footnote{E-mail: \texttt{qiuyu@ustc.edu.cn}}
{\large }
\bigskip
\\
  {\small \it $^a$Department of Modern Physics, University of Science
    and}
\\
{ {\small \it  Technology  of China, Hefei, Anhui 230026, China.}}
\\
{\small \it $^b$INFN, Sezione di Torino, Via P. Giuria 1, I-10125
Torino, Italy} }

\maketitle \vskip 12mm

\begin{abstract}
We propose a seesaw scenario that possible corrections to the
tribimaximal pattern of lepton mixing are due to the small phase
splitting of the right-handed neutrino mass matrix. we show that the
small deviations can be expressed analytically in terms of two
splitting parameters($\delta_1$ and $\delta_2$) in the leading
order. The solar mixing angle $\theta_{12}$ favors a relatively
smaller value compared to zero order value ($35.3^\circ$), and the
Dirac type CP phase $\delta$ chooses a nearly maximal one. The two
Majorana type CP phases $\rho$ and $\sigma$ turn out to be a nearly
linear dependence. Also a normal hierarchy neutrino mass spectrum is
favored due to the stability of perturbation calculations.
\end{abstract}

\vspace{3mm}

\vskip 12mm

{\small PACS numbers: 14.60.Pq, 14.60.St, 11.30Hv

}

\vskip 5cm {\large \textbf{ 
}} \vfill \eject
\baselineskip=0.30in
\renewcommand{\theequation}{\arabic{section}.%
\arabic{equation}} \renewcommand{\thesection}{\Roman{section}}
\makeatletter
\@addtoreset{equation}{section} \makeatother

\section{Introduction}
Due to the out-standing achievements in the neutrino oscillation
experiments\cite{sno1,sno2,sno3,sno4,sno5,Kamiokande1,Kamiokande2,Kamiokande3,SK1,SK2,SK3,kldet1,kldet2,kldet3,CHOOZ,K2K1,K2K2,MINOS1},
we have convincing evidences on neutrino mass and lepton mixing. In
the weak basis that the charged-lepton mass matrix is diagonal, real
and positive\,, we assume massive neutrinos to be Majorana particles
and parametrize\cite{PDG,SV} the lepton mixing $V$ as
\begin{equation}\label{sp}
V = \left( \matrix{ c^{}_{12}c^{}_{13} & s^{}_{12}c ^{}_{13} &
s^{}_{13} e^{-i\delta} \cr -s^{}_{12}c^{}_{23}
-c^{}_{12}s^{}_{23}s^{}_{13} e^{i\delta} & c^{}_{12}c^{}_{23}
-s^{}_{12}s^{}_{23}s^{}_{13} e^{i\delta} & s^{}_{23}c^{}_{13} \cr
s^{}_{12}s^{}_{23} -c^{}_{12}c^{}_{23}s^{}_{13} e^{i\delta} &
-c^{}_{12}s^{}_{23} -s^{}_{12}c^{}_{23}s^{}_{13} e^{i\delta} &
c^{}_{23}c^{}_{13} } \right) \left ( \matrix{e^{i\rho } & 0 & 0 \cr
0 & e^{i\sigma} & 0 \cr 0 & 0 & 1 \cr} \right ) \,.
\end{equation}
Now the global fit\cite{SV,GF1,GF2} of current experimental dada
yields $\sin^2\theta_{12} \sim 0.304^{+0.022}_{-0.016}$\,,
$\sin^2\theta_{23} \sim 0.50^{+0.07}_{-0.06}$\,,$\sin^2\theta_{13}
\leq 0.016$\, at 1 $\sigma$. So the so-called tribimaximal
mixing\cite{TB1,TB2,TB3,TB4} pattern is an excellent approximation
for these physical values, which appears as the form of
\begin{equation}\label{TB}
V^{}_0 = \left( \matrix{ 2/\sqrt{6} & 1/\sqrt{3} & 0 \cr -1/\sqrt{6}
& 1/\sqrt{3} & 1/\sqrt{2} \cr 1/\sqrt{6} & -1/\sqrt{3} & 1/\sqrt{2}
} \right)  \; .
\end{equation}
It corresponds to $\sin^{2}\theta_{12}=1/3$\,,
$\sin^{2}\theta_{23}=1/2$\, and $\sin^{2}\theta_{13}=0$ in standard
parametrization. This simple form with small integers motivates
neutrino theorists to consider some underlying structures. The most
intriguing one is flavor symmetry among generations, especially for
discrete groups\cite{FS1,FS2} such as $A_4$\,, $S_3$\,, $Z_2$, which
can give predictable values or/and relations of mixing parameters.
In general, nearly tribimaximal mixing is also allowed, so small
corrections to the standard form in (\ref{TB}) are interesting and
necessary. Many sources can give these corrections such as charged
lepton sector contributions\cite{TB51,TB52,TB53} and renormalization
group effects\cite{TB61,TB62,TB63}.\\
On the other way, seesaw mechanism\cite{SS,SS1,SS2,SS3,SS4} can
naturally explain the smallness of three left-handed neutrino
masses. In the simplest type I framework, the effective Majorana
mass matrix of neutrinos $M_{\nu}$ is related to the Dirac mass
matrix $M_{D}$ and the heavy right-handed Majorana mass matrix
$M_{R}$ by the relation
\begin{equation}\label{ss1}
M_{\nu}\,{\simeq}\,M_{D}M_{R}^{-1}M_{D}^{T}.
\end{equation}
To explore the structure of the seesaw formula, the effective mixing
can generally be derived from both $M_{D}$ and $M_{R}$. One of them
can be chosen to be diagonal or even identity matrix for
simplifications. One kind of models in ref.\cite{GL1,GL2,GL3} takes
a diagonal $M_{D}$ for instance. Meanwhile, another kind of seesaw
models\cite{KX1,KX2} starts from a unit form of $M_{R}$.\\
Now for the small corrections of tribimaximal mixing, another
source\cite{KX1,KX2} in realistic seesaw models attracts our
attentions. The leading order mixing matrix is totally derived from
the Dirac mass matrix $M_{D}$, and $M_{R}$ has the simplest form of
$M_{0}I$ in the symmetry limit ($M_{0}$ being a common mass scale
and $I$ the identity matrix)\,. The departure of $M_{R}$ from the
unit matrix gives small corrections to the effective mixing matrix.
There are two methods in $M_{R}$ to break the unit form, one is the
nondegenerate masses, another is the un-degenerate phases: the phase
breaking. In this letter, we want to discuss the second case, we
consider small phase splitting for complex matrix $M_{R}$, and
maintain the degeneracy of the right-handed neutrino masses. We find
that besides small corrections of the three mixing angles, we get
nontrivial CP violation phases.\\
The remaining part of this paper is organized as follows. In section
II, we talk about the realistic models from phase breaking, giving
the predicted values of mixing parameters. In section III, we do
some numerical analysis, displaying the correlations of these
parameters. We conclude our topic and give some remarks in section
IV. In the Appendix, we give our calculations from perturbation
approximations.

\section{realistic models}
In our model, we adopt the basis started in\cite{KX1,KX2}. we think
$M_{D}$ can be diagonalized by tribimaximal mixing matrix in the
manner of $V^{+}_0M_{D}V^{*}_0{=}Diag\{x,y,z\}$ which is constrained
by a discrete flavor symmetry(such as $S_3$ symmetry). $(x,y,z)$ are
positive mass parameters. For $M_{R}$, we take the form of $M_{0}I$
in the symmetry limit, obviously it is invariant under the symmetry.
Then we have
\begin{eqnarray}\label{seesaw1}
M_{\nu}\,{=}\,M_{D}M_{R}^{-1}M_{D}^{T} {=} V^{}_0 \left( \matrix{ x
& 0 & 0 \cr 0 & y & 0 \cr 0 & 0 & z } \right) V^{T}_0
\frac{1}{M^{}_{0}} V^{}_0 \left( \matrix{ x & 0 & 0 \cr 0 & y & 0
\cr 0 & 0 & z } \right) V^{T}_0
\nonumber\\
{=} V^{}_0 \frac{1}{M^{}_{0}} \left( \matrix{ x^{2} & 0 & 0 \cr 0 &
y^{2} & 0 \cr 0 & 0 & z^{2} } \right) V^{T}_0 \equiv V^{}_0
M^{(0)}_{\nu} V^{T}_0\,.
\end{eqnarray}
In the following, we want to give small phase splitting to $M_{R}$,
and derive the light neutrino mass spectra and small corrections for
lepton mixing parameters compared with the leading values of
$V^{}_0$ including CP violating phases\,.

\subsection{Scenario A}
In this scenario, we give ${M_{R}}_{1,1}$ a small phase compared to
the others, then $M_{R}$ take the form of
\begin{equation}
M_{R} =  M_{0}\left( \matrix{ e^{-i\delta_{1}} & 0 & 0 \cr 0 & 1 & 0
\cr 0 & 0 & 1 } \right)  \; .
\end{equation}
Similar to (\ref{seesaw1}), the effective neutrino mass matrix
$M_{\nu}$ turns out to be
\begin{eqnarray}
M_{\nu}\,{=}\,M_{D}M_{R}^{-1}M_{D}^{T} {=} V^{}_0 \left( \matrix{ x
& 0 & 0 \cr 0 & y & 0 \cr 0 & 0 & z } \right) V^{T}_0
\frac{1}{M^{}_{0}}\left( \matrix{ e^{i\delta_{1}} & 0 & 0 \cr 0 & 1
& 0 \cr 0 & 0 & 1 } \right) V^{}_0 \left( \matrix{ x & 0 & 0 \cr 0 &
y & 0 \cr 0 & 0 & z } \right) V^{T}_0
\nonumber\\
{=} V^{}_0 \frac{z^{2}}{3M^{}_{0}} \left( \matrix{
(3+2\epsilon_{1})\omega^{2} \eta^{2} & \sqrt{2}\omega \eta^{2} & 0
\cr \sqrt{2}\omega \eta^{2} & (3+\epsilon_{1})\eta^{2} & 0 \cr 0 & 0
& 3 } \right) V^{T}_0 \equiv V^{}_0 M^{(1)}_{\nu} V^{T}_0\,,
\end{eqnarray}
with $\omega \equiv x/y $\,, $\eta \equiv y /z $ and $\epsilon_{1}
\equiv e^{i\delta_{1}}-1$\,. The corrections to $V^{}_{0}$ come from
the diagonalization of matrix $M^{(1)}_{\nu}$. Assuming
$M^{(1)}_{\nu}\equiv V^{}_{1}\overline{M}^{}_{\nu}V^{T}_{1}$\,,
where $\overline{M}^{}_{\nu}=Diag\{m_{1},m_{2},m_{3}\}$ with $m_{i}
(i=1,2,3)$ being the Majorana neutrino masses. When we parameterize
$V^{}_{1}$ as
\begin{equation}
V^{}_{1} = \left( \matrix{ \cos\alpha & \sin\alpha e^{-i\beta} & 0
\cr -\sin\alpha e^{i\beta} & \cos\alpha & 0 \cr 0 & 0 & 1 } \right)
\left ( \matrix{e^{i\sigma_{1}} & 0 & 0 \cr 0 & e^{i\sigma_{2}} & 0
\cr 0 & 0 & e^{i\sigma_{3}} \cr} \right ) \; ,
\end{equation}
we can derive these mixing parameters and masses by the relation of
$\overline{M}^{}_{\nu}=V^{+}_{1}M^{(1)}_{\nu}V^{*}_{1}$ in terms of
$\epsilon_{1}$\,, $\omega$\,, $\eta$\,, $z$\, and $M^{}_{0}$\,. In
our calculation, we view $\epsilon_{1}$ (or equivalently
$\delta_{1}$) as a small parameter, and give results with the
leading terms of the power series of $\delta_{1}$\,, such as
$\epsilon_{1} \equiv e^{i\delta_{1}}-1 \simeq -\delta^{2}_{1}/2 + i
\delta_{1}$\,. By the perturbation calculations, we derive the
relation of $\cos\beta \simeq 1/6 \sin\beta \delta_{1}$\,,
immediately we have $\cot\beta \simeq \delta_{1}/6$\,. Other
parameters are listed below
\begin{equation}
\tan2\alpha \simeq -\frac{2\sqrt{2}}{3} \frac{\omega}{1+\omega^{2}}
\,\delta_{1},
\end{equation}
\begin{eqnarray}
m_{1} \simeq \frac{z^{2}\omega^{2}\eta^{2}}{M^{}_{0}}
\{1-\frac{1}{9} \delta^{2}_{1} +
\frac{2}{9}\frac{1}{1+\omega^{2}}\delta^{2}_{1}\} \simeq
\frac{z^{2}\omega^{2}\eta^{2}}{M^{}_{0}}(1 + \frac{1}{9}
\delta^{2}_{1})\,,
\nonumber\\
m_{2} \simeq \frac{z^{2}\eta^{2}}{M^{}_{0}} \{1-\frac{1}{9}
\delta^{2}_{1} + \frac{2}{9}\frac{\omega^{2}}{1+\omega^{2}}
\delta^{2}_{1} \} \simeq
\frac{z^{2}\eta^{2}}{M^{}_{0}}(1-\frac{1}{9} \delta^{2}_{1})\,,
\end{eqnarray}
and
\begin{equation}
m_{3} = \frac{z^{2}}{M^{}_{0}}\;,\;\tan2\sigma_{1} \simeq
\frac{2}{3}\delta_{1}\;,\;  \tan2\sigma_{2} \simeq
\frac{1}{3}\delta_{1}\;,\; \tan2\sigma_{3} = 0\,.
\end{equation}
For the second expressions of $m_{1}$ and $m_{2}$\,, we omit the
contributions of higher powers of $\omega$\,. Combining $V^{}_{1}$
together with $V^{}_{0}$\, and making a rephasing transformation, we
can get the standard expression of neutrino mixing matrix $V$ as
eq.(\ref{sp})
\begin{eqnarray}
V \equiv V^{}_{0}V^{}_{1} = \left( \matrix{ \frac{2}{\sqrt{6}} &
\frac{1}{\sqrt{3}} & 0 \cr -\frac{1}{\sqrt{6}} & \frac{1}{\sqrt{3}}
& \frac{1}{\sqrt{2}} \cr \frac{1}{\sqrt{6}} & -\frac{1}{\sqrt{3}} &
\frac{1}{\sqrt{2}} } \right) \left( \matrix{ \cos\alpha & \sin\alpha
e^{-i\beta} & 0 \cr -\sin\alpha e^{i\beta} & \cos\alpha & 0 \cr 0 &
0 & 1 } \right) \left ( \matrix{e^{i\sigma_{1}} & 0 & 0 \cr 0 &
e^{i\sigma_{2}} & 0 \cr 0 & 0 & e^{i\sigma_{3}} \cr} \right )\,
\nonumber\\
\equiv \left ( \matrix{e^{i\rho_{1}} & 0 & 0 \cr 0 & e^{i\rho_{2}} &
0 \cr 0 & 0 & e^{i\rho_{3}} \cr} \right ) \left( \matrix{
\cos\theta_{12} & \sin\theta_{12} & 0 \cr
-\frac{1}{\sqrt{2}}\sin\theta_{12} &
\frac{1}{\sqrt{2}}\cos\theta_{12} & \frac{1}{\sqrt{2}} \cr
\frac{1}{\sqrt{2}}\sin\theta_{12} &
-\frac{1}{\sqrt{2}}\cos\theta_{12} & \frac{1}{\sqrt{2}} } \right)
\left ( \matrix{e^{i\rho} & 0 & 0 \cr 0 & e^{i\sigma} & 0 \cr 0 & 0
& 1 \cr} \right )\,,
\end{eqnarray}
where $\rho_{i} (i=1,2,3)$ can be rotated away by redefining the
phases of three charged-lepton fields. Then the standard parameters
appear as
\begin{equation}
\sin\theta_{12} \simeq \frac{1}{\sqrt{3}}\{
1-\frac{\omega}{1+\omega^{2}}
(1-\frac{\omega}{1+\omega^{2}})\frac{1}{9}\delta^{2}_{1}\}\,,
\end{equation}
and two standard Majorana phases are
\begin{eqnarray}
\tan2\rho \simeq \frac{2}{3}(1-\frac{2\,
\omega}{1+\omega^{2}})\delta_{1}\,,
\nonumber\\
\tan2\sigma \simeq \frac{1}{3}(1-\frac{2\,
\omega}{1+\omega^{2}})\delta_{1}\,.
\end{eqnarray}
In this scenario, we get a small deviation from tribimaximal mixing
for $\theta_{12}$\,, and two Majorana phases, but $\theta_{13}$ and
$\theta_{23}$ remain unchanged. Because of the vanishing value of
$\theta_{13}$\,, we cannot obtain the information of Dirac CP phase.
Contrarily if we give ${M_{R}}_{2,2}$ another small phase compared
to ${M_{R}}_{3,3}$, we can obtain a non-vanishing $\theta_{13}$ and
then a non-trivial Dirac CP violation phase.
\subsection{Scenario B}
As discussed in the end of last section, we take a general form of
$M_{R}$ as
\begin{equation}\label{mr}
M_{R} =  M_{0}\left( \matrix{ e^{-i\delta_{1}} & 0 & 0 \cr 0 &
e^{-i\delta_{2}} & 0 \cr 0 & 0 & 1 } \right)  \; .
\end{equation}
Then we will repeat the same procedure as before. Firstly the
effective neutrino mass matrix are
\begin{eqnarray}\label{mb}
M_{\nu}\,{=}\,M_{D}M_{R}^{-1}M_{D}^{T} {=} V^{}_0 \left( \matrix{ x
& 0 & 0 \cr 0 & y & 0 \cr 0 & 0 & z } \right) V^{T}_0
\frac{1}{M^{}_{0}}\left( \matrix{ e^{i\delta_{1}} & 0 & 0 \cr 0 &
e^{i\delta_{2}} & 0 \cr 0 & 0 & 1 } \right) V^{}_0 \left( \matrix{ x
& 0 & 0 \cr 0 & y & 0 \cr 0 & 0 & z } \right) V^{T}_0
\nonumber\\
{=} V^{}_0 \frac{z^{2}}{6M^{}_{0}} \left( \matrix{
(6+4\epsilon_{1}+\epsilon_{2})\omega^{2} \eta^{2} &
(2\epsilon_{1}-\epsilon_{2})\sqrt{2}\omega \eta^{2} &
-\sqrt{3}\epsilon_{2}\omega \eta \cr
(2\epsilon_{1}-\epsilon_{2})\sqrt{2}\omega \eta^{2} &
(6+2\epsilon_{1}+2\epsilon_{2})\eta^{2} & \sqrt{6}\epsilon_{2}\eta
\cr -\sqrt{3}\epsilon_{2}\omega \eta & \sqrt{6}\epsilon_{2}\eta &
6+3\epsilon_{2}} \right) V^{T}_0 \equiv V^{}_0 M^{(2)}_{\nu}
V^{T}_0\,.
\end{eqnarray}
Then we define $M^{(2)}_{\nu} \equiv V^{}_{2} \overline{M}^{}_{\nu}
V^{T}_{2}$\,, with $V^{}_{2}$ is an unitary matrix parameterized as
\begin{equation}\label{par2}
V^{}_{2} = \left( \matrix{ D_{1} & \alpha_{3} & \alpha_{2} \cr
-\alpha^{*}_{3} & D_{2} & \alpha_{1} \cr -\alpha^{*}_{2} &
-\alpha^{*}_{1} & D_{3} } \right) \left ( \matrix{e^{i\sigma_{1}} &
0 & 0 \cr 0 & e^{i\sigma_{2}} & 0 \cr 0 & 0 & e^{i\sigma_{3}} \cr}
\right ) \; ,
\end{equation}
where $\alpha_{j}\equiv \sin\theta_{j}e^{-i\phi_{j}}$(for $j=1,2,3$)
, $D_{1}=\sqrt{1-|\alpha_{2}|^{2}-|\alpha_{3}|^{2}}$ and similar
definitions for $D_{2}$ and $D_{3}$\,. By using
$\overline{M}^{}_{\nu}=V^{+}_{2}M^{(2)}_{\nu}V^{*}_{2}$\,, we arrive
at the approximate expressions for $\theta_{i}$\, and $\phi_{i}$\,
via perturbation calculations.(we will give a careful calculations
of perturbation approximations in the Appendix.)
\begin{equation}\label{sinsin}
\sin\theta_{3}\sin\phi_{3} \simeq - \frac{\sqrt{2}}{6}\omega
(2\delta_{1}-\delta_{2})\;, \sin\theta_{2}\sin\phi_{2} \simeq
\frac{\sqrt{3}}{6}\omega\eta\delta_{2}\;, \sin\theta_{1}\sin\phi_{1}
\simeq -\frac{\sqrt{6}}{6}\eta\delta_{2}\;;
\end{equation}
and
\begin{eqnarray}\label{sincos}
\sin\theta_{3}\cos\phi_{3} \simeq -\frac{\sqrt{2}}{18}\omega(
\delta^{2}_{1}-\delta_{1}\delta_{2}+ \delta^{2}_{2} )\,,
\nonumber\\
\sin\theta_{2}\cos\phi_{2} \simeq
\frac{\sqrt{3}}{18}\,\omega\,\eta\,\delta_{2}
(2\delta_{1}-\delta_{2})\,,
\nonumber\\
\sin\theta_{1}\cos\phi_{1} \simeq -\frac{\sqrt{6}}{36}
\eta(\omega^{2}-\eta^{2}) \delta_{2}(2\delta_{1}-\delta_{2})\,.
\end{eqnarray}
And the masses are
\begin{eqnarray}\label{mi}
m_{1} \simeq \frac{z^{2}\omega^{2}\eta^{2}}{M^{}_{0}}(1 +
\frac{1}{9} \delta^{2}_{1} - \frac{1}{9}\delta_{1}\delta_{2} +
\frac{5}{72} \delta^{2}_{2})\,,
\nonumber\\
m_{2} \simeq \frac{z^{2}\eta^{2}}{M^{}_{0}}(1 - \frac{1}{9}
\delta^{2}_{1} + \frac{1}{9}\delta_{1}\delta_{2} + \frac{1}{18}
\delta^{2}_{2})\,,
\nonumber\\
m_{3} \simeq \frac{z^{2}}{M^{}_{0}}(1 - \frac{1}{8}
\delta^{2}_{2})\;.
\end{eqnarray}
Finally we give the phase-angles $\sigma_{i}$ in eq.(\ref{par2})
\begin{equation}
\tan2\sigma_{1} \simeq \frac{1}{6}(4\delta_{1}+\delta_{2})\;,\;
\tan2\sigma_{2} \simeq \frac{1}{3}(\delta_{1}+\delta_{2})\;,\;
\tan2\sigma_{3} \simeq \frac{1}{2}\delta_{2}\;.
\end{equation}
A normal hierarchy mass spectrum(small values of $\omega$ and
$\eta$) is favored in this scenario due to the stability of
perturbation calculation(see the Appendix). So for the analytical
expressions in (\ref{sinsin}) (\ref{sincos}) and (\ref{mi}), we
express them with the leading terms of the power series of $\omega$
and $\eta$. More detailed expressions and discussions can be found
in the Appendix. \\
Now we want to exhibit the mixing matrix $V =V_{0}V_{2}$ with the
standard parametrization as in eq.(\ref{sp}). By a rephasing
transformation, we get
\begin{eqnarray}
&&V = \left( \matrix{ \frac{2}{\sqrt{6}} & \frac{1}{\sqrt{3}} & 0
\cr -\frac{1}{\sqrt{6}} & \frac{1}{\sqrt{3}} & \frac{1}{\sqrt{2}}
\cr \frac{1}{\sqrt{6}} & -\frac{1}{\sqrt{3}} & \frac{1}{\sqrt{2}} }
\right) \left( \matrix{ D_{1} & \alpha_{3} & \alpha_{2} \cr
-\alpha^{*}_{3} & D_{2} & \alpha_{1} \cr -\alpha^{*}_{2} &
-\alpha^{*}_{1} & D_{3} } \right) \left ( \matrix{e^{i\sigma_{1}} &
0 & 0 \cr 0 & e^{i\sigma_{2}} & 0 \cr 0 & 0 & e^{i\sigma_{3}} \cr}
\right ) \equiv \left ( \matrix{e^{i\rho_{1}} & 0 & 0 \cr 0 &
e^{i\rho_{2}} & 0 \cr 0 & 0 & e^{i\rho_{3}} \cr} \right )
\nonumber\\
&&\left( \matrix{ c^{}_{12}c^{}_{13} & s^{}_{12}c ^{}_{13} &
s^{}_{13} e^{-i\delta} \cr -s^{}_{12}c^{}_{23}
-c^{}_{12}s^{}_{23}s^{}_{13} e^{i\delta} & c^{}_{12}c^{}_{23}
-s^{}_{12}s^{}_{23}s^{}_{13} e^{i\delta} & s^{}_{23}c^{}_{13} \cr
s^{}_{12}s^{}_{23} -c^{}_{12}c^{}_{23}s^{}_{13} e^{i\delta} &
-c^{}_{12}s^{}_{23} -s^{}_{12}c^{}_{23}s^{}_{13} e^{i\delta} &
c^{}_{23}c^{}_{13} } \right) \left ( \matrix{e^{i\rho } & 0 & 0 \cr
0 & e^{i\sigma} & 0 \cr 0 & 0 & 1 \cr} \right ) \;.
\end{eqnarray}
By a lengthy but straightforward calculation, we can get the
standard predictions. The Dirac type CP phase is predicted as
\begin{equation}{\label{delta}}
\cos\delta \simeq
\frac{5}{6}\frac{\omega}{1-\omega}(2\delta_{1}-\delta_{2})\,,
\end{equation}
and the Majorana CP phases are
\begin{eqnarray}{\label{rhosigma}}
\tan2\rho \simeq \frac{1}{3}(1-2\,\omega)(2\delta_{1}-\delta_{2})\,,
\nonumber\\
\tan2\sigma \simeq \frac{1}{6}(1-2\,
\omega)(2\delta_{1}-\delta_{2})\,.
\end{eqnarray}
About the mixing angles, we can obtain a small departure from the
tribimaximal mixing in the manner of
\begin{eqnarray}{\label{corb}}
\sin\theta_{13} \simeq \frac{\sqrt{2}}{6}\{\frac{\eta}{1+\eta^{2}}-
\frac{\omega\eta}{1+\omega^{2}\eta^{2}}\}|\delta_2|\,,
\nonumber\\
\tan\theta_{12} \simeq \frac{\sqrt{2}}{2}\{1-\frac{1}{6}\omega
(\delta^{2}_{1}-\delta_{1}\delta_{2}+\delta^{2}_{2})\}\,,
\nonumber\\
\tan\theta_{23} \simeq
1-\frac{1}{9}\,\omega\eta\,\delta_{2}(2\delta_{1}-\delta_{2})\,.
\end{eqnarray}
Immediately, when $\delta_2\rightarrow 0$, the predicted values
return to the corresponding ones in \emph{Scenario A} as we expect
to, and the Dirac phase $\delta$ is undetermined due to zero of
$\sin\theta_{13}$. The small parameter $\delta_2$ is responsible for
the non-vanishing $\theta_{13}$ and non-maximality of $\theta_{23}$,
and the mixing angle $\theta_{12}$ depends on both $\delta_1$ and
$\delta_2$. In this scenario, we predict a nearly maximal Dirac
phase, so the rephasing invariant Jarlskog parameter $J$\cite{J,Wu}
is determined only by the three mixing angles in leading order
\begin{equation}
J \simeq \frac{\sqrt{2}}{6}\sin\theta_{13} \simeq
\frac{1}{18}\eta(1-\omega)|\delta_2|\,.
\end{equation}
Indeed, the most general Dirac mass matrix $M_{D}$ can be
diagonalized by two distinct unitary matrices $V_{L}$ and $V_{R}$,
so if we identify $V_{L}$ as the exact tribimaximal form, the high
order corrections come from the contribution of $N_{R} \equiv
V^{+}_{R}M^{-1}_{R}V^{*}_{R}$, which is a unitary symmetric matrix
if the mass parameters in $M_{R}$ are degenerated. This case has
been discussed for the (light) effective Majorona mass matrix with
exact degenerate masses\cite{deg1,deg2}. They found there are only
two mixing angles and one CP phase in the parametrization of mixing
matrix, but it does not include the case with the same CP parity as
a limit. In our scenarios, we want to get higher order corrections
from the diagonalization of $N_{R}\,$, so we are only interested in
the nearly unit form of $N_{R}\,$. It is convenient to parameterized
as $N_{R} \equiv V^{T}_{N}D_{\delta_{i}}V_{N}$ where $V_{N}$ is a
real unitary (orthogonal) matrix and $D_{\delta_{i}}$ is a diagonal
unitary matrix with small $\delta_{i}$ denoting the small departure
from unit form of $N_{R}$. In the beginning of our calculation, we
have identified $V_{N}$ with the explicit form of $V_{0}\,$ for
simplicity. Although this is only a special example, but it is
enough to reveal the property of our scenarios and correlations of
these corrections. In next section, we will show them numerically.

\section{numerical analysis}
\emph{Scenario B} returns to \emph{Scenario A} when
$\delta_2\rightarrow 0$\,, so we can only concern the general case.
To do numerical analysis, we will include higher order contributions
of $\omega$ and $\eta$\,as in ((5.2), (5.7)--(5.9)). Firstly, the
correlations of $\omega$ and $\eta$\, come form the ratio of $\Delta
m^2_{Sol}$ and $\Delta m^2_{Atm}$ by the relation of
\begin{equation}
\frac{\Delta m^2_{Sol}}{\Delta m^2_{Atm}}\simeq
\frac{\eta^4(1-\omega^4)}{1-\eta^4}.
\end{equation}
Using the numerical results in \cite{GF1,GF2} which reveal that
$\Delta m^2_{Sol} = 7.65^{+0.23}_{-0.20}\times10^{-5} eV^2$ and
$\Delta m^2_{Atm} = 2.40^{+0.12}_{-0.11}\times10^{-3} eV^2$ at 1
$\sigma$ , we can obtain the allowed region of $\omega$ versus
$\eta$ in FIG.1. The mass spectrum changes from hierarchy region to
nearly degenerate region as $\omega$ grows. And $\eta$ takes a
nearly fixed value in the hierarchy region. As discussed in the
Appendix, only the normal hierarchy mass spectrum is valid in our
scenarios, so we restrict
the parameter $\eta\leq 0.5$ in our analysis.\\
The departure of mixing matrix from the standard tribimaximal mixing
depends on the two small parameters $\delta_1$ and $\delta_2$. If we
change the values of $\delta_i$\,, we can get the correlation
relations between mixing parameters(mixing angles and CP phases)
defined in (\ref{sp}). Scanning $\delta_i$ within a reasonable
range(\,[\,--\,0.5\,,\,0.5\,]), we give five pictures FIG.2--FIG.6
as typical examples. The angles and phases are measured in degrees.
Some comments are listed below. \\
$\bullet$ For the relation of between $\theta_{12}$ and
$\theta_{13}$ in FIG.2\,, we can see that the correction of
$\theta_{12}$ is only at the left side, which has been revealed in
(\ref{corb}) for leading corrections, indicating that a relatively
small $\theta_{12}$ is favored, which is in accordance with the best
fit point ($33.2^\circ$) of global analysis\cite{GF}. The magnitude
of $\theta_{13}$ is too small ($<2.5^\circ$) to be measured in the
near future. The sensitivity of the proposed reactor neutrino
experiments to $\theta_{13}$ is at the level of
$\theta_{13}\sim3^\circ$($\sin^2{2\theta_{13}}\sim0.01$)\cite{SV,daya}.
Moreover, a larger deviation of $\theta_{12}$ from $35.3^\circ$
($\sin^{-1}{\sqrt{\frac{1}{3}}}$) implies a more
stringent constraint on $\theta_{13}$, including a lower bound.\\
$\bullet$ In FIG.3\,, the deviation of $\theta_{23}$ can extend to
both side, but with an unsymmetrical shape. The magnitude of this
deviation is less than $1^\circ$. A larger $\theta_{13}$ allows a
smaller region of $\theta_{23}$. Oppositely, a larger deviation of
$\theta_{23}$ from maximality implies a smaller range for $\theta_{13}$. \\
$\bullet$ The Dirac type CP phase $\delta$ favors a nearly maximal
value which can be understood in (\ref{delta}) and in FIG.4. The
relatively broad width results from the large range of
$(2\delta_1-\delta_2)$. The figure shows that the more deviation of
$\delta$ from maximality, the more stringent upper bound
$\theta_{13}$ suffers. The Jarlskog parameter $J$ can reach the
order of $10^{-2}$ at most, which is
limited by the relatively small value of $\theta_{13}$\,.\\
$\bullet$ Just as revealed in FIG.5, the two Majorana CP phases
$\rho$ and $\sigma$ have distributions among
[\,$-180^\circ$,\,$180^\circ$\,] peaked at zero, which is consistent
with leading predictions of (\ref{rhosigma}). Also, FIG.6 shows that
$\rho$ and $\sigma$ are strongly correlated with each other, having
a nearly linear dependence. The slope of $\rho$ versus $\sigma$
defined in (\ref{rhosigma}) approximates to $0.5$\,, and the
corresponding one in FIG.6 lies between $0.75\sim1.0$\,, where the
higher order contributions of $\omega$ and $\eta$ have been
included.
\section{Conclusions}
To conclude, we discuss possible corrections to the tribimaximal
pattern of lepton mixing in a simple seesaw model, which is the
results of phase breaking of right-handed neutrino matrix. We
consider small phase splitting for complex matrix $M_{R}$, but
maintain the degeneracy of the right-handed neutrino masses. The
breaking of the unit form of $M_{R}$ in the seesaw model gives small
corrections to the zero order form of the mixing matrix. As revealed
in (\ref{corb}), the corrections of $\theta_{13}$\,,
$\theta_{12}$\,, and $\theta_{23}$ are of order $O(\delta_i)$\,,
$O(\delta_i^{2})$\,, and $O(\delta_i^{2})$\, respectively.
$\theta_{12}$ favors a smaller value compared to the zero order
value of $35.3^\circ$\,. The Dirac type CP phase $\delta$ is likely
to have a nearly maximal value. For the two Majorana CP phases
$\rho$ and $\sigma$, they have a nearly linear correlation.\\
A normal hierarchy mass spectrum is favored in our scenarios,
leading to insignificant renormalization group
effects\cite{TB61,TB62,TB63,KX1,KX2} of the mixing matrix. The exact
mass degeneracy of the righthanded neutrinos forbids CP violation in
the lepton-number-violating decays\cite{LFV}, so there is no thermal
leptogenesis\cite{FY}. Generally, we should include both effects
mentioned in the introduction: the nondegenerate masses and the
nondegenerate phases. But our discussions are valid when the mass
degeneracy breaking is much smaller than the latter one. The general
cases including both (and the corresponding leptogenesis) are
certainly interesting and
need further discussions.\\
There are other interesting aspects on the corrections to
tribimaximal mixing based on either different
assumptions\cite{HE,Tanimoto} or general
parameterizations\cite{PA1,PA2,PA3}. They can only be distinguished
by precision measurements in future reactor and accelerator neutrino
experiments\cite{SV,DCH,daya}.

\acknowledgments{The authors are grateful to B.L. Chen for the help
of numerical analysis. This work is supported in part by the
National Natural Science Foundation of China under grant number
90203002. The author (YFL) would like to thank the Department of
Theoretical Physics of the University of Torino for hospitality and
support.}

\section{Appendix: Perturbation Calculations for Diagonalizing the neutrino
mass matrix} When we want to diagonalize the matrix of
$M^{(2)}_{\nu}$ in (\ref{mb}), we refer to the assumption that
parameters $\delta_i$ are small enough to do perturbation
calculations. In the relation of
$\overline{M}^{}_{\nu}=V^{+}_{2}M^{(2)}_{\nu}V^{*}_{2}$\,,we think
equations are realized in each order for the power series of
$\delta_i$. Then when we put the approximation to the order of
$\delta_i$\, ($O(\delta_1,\delta_2)$), we can get three equations by
the relations of three off-diagonal elements
\begin{eqnarray}\label{equation1}
i\sqrt{2}(2\delta_{1}-\delta_{2})\omega\eta^{2}+\sin\theta_{3}
(\cos\phi_{3}+i\sin\phi_{3})6\omega^{2}\eta^{2}-\sin\theta_{3}
(\cos\phi_{3}-i\sin\phi_{3})6\eta^{2} = 0 \,,
\nonumber\\
-i\sqrt{3}\delta_{2}\omega\eta + \sin\theta_{2}
(\cos\phi_{2}+i\sin\phi_{2})6\omega^{2}\eta^{2} - \sin\theta_{2}
(\cos\phi_{2}-i\sin\phi_{2})6 = 0 \,,
\nonumber\\
i\sqrt{6}\delta_{2}\eta + \sin\theta_{1}
(\cos\phi_{1}+i\sin\phi_{1})6\eta^{2}-\sin\theta_{1}
(\cos\phi_{1}-i\sin\phi_{1})6 = 0 \,.
\end{eqnarray}
Then we get the results as
\begin{eqnarray}\label{sinsin2}
\sin\theta_{3}\sin\phi_{3} \simeq
-\frac{\sqrt{2}\omega}{6(1+\omega^{2})} (2\delta_{1}-\delta_{2})\;,
\nonumber\\
\sin\theta_{2}\sin\phi_{2} \simeq
\frac{\sqrt{3}\omega\eta}{6(1+\omega^{2}\eta^{2})}\delta_{2}\;,
\nonumber\\
\sin\theta_{1}\sin\phi_{1} \simeq
-\frac{\sqrt{6}\eta}{6(1+\eta^{2})}\delta_{2}\;,
\end{eqnarray}
and
\begin{equation}\label{sincos2}
\sin\theta_{3}\cos\phi_{3} = \sin\theta_{2}\cos\phi_{2} =
\sin\theta_{2}\cos\phi_{2} = 0\,.
\end{equation}
Which indicate maximality for $\phi_i$($\cos\phi_i=0$). We want to
know the departure from maximality in order of
$O(\delta_1,\delta_2)$\,, so terms in order of $O(\delta^{2}_1,
\delta_1\delta_2, \delta^2_2)$\, must be considered in equations of
(\ref{equation1}). Up to this order those equations have the form of
\begin{eqnarray}
i\sqrt{2}(2\delta_{1}-\delta_{2})\omega\eta^{2}+i\sin\theta_{3}
\sin\phi_{3}6\eta^{2}(1+\omega^{2})+\sin\theta_{3}
\cos\phi_{3}6\eta^{2}(\omega^{2}-1)
\nonumber\\
+\sqrt{2}(-\delta^{2}_{1}+\frac{\delta^{2}_{2}}{2})\omega\eta^{2}
-\sin\theta_{3}\sin\phi_{3}\eta^{2}[(4\delta_{1}+\delta_{2})\omega^{2}+
(2\delta_{1}+2\delta_{2})]
\nonumber\\
-6\sin\theta_{1}\sin\phi_{1}\sin\theta_{2}\sin\phi_{2} -
\sqrt{6}\sin\theta_{2}\sin\phi_{2}\delta_{2}\eta +
\sqrt{3}\sin\theta_{1}\sin\phi_{1}\delta_{2}\omega\eta = 0 \,,
\end{eqnarray}
\begin{eqnarray}
-i\sqrt{3}\delta_{2}\omega\eta + i\sin\theta_{2}
\sin\phi_{2}6(1+\omega^{2}\eta^{2})+\sin\theta_{2}
\cos\phi_{2}6(\omega^{2}\eta^{2}-1)
+\frac{\sqrt{3}}{2}\delta^{2}_{2}\omega\eta
\nonumber\\
-\sin\theta_{2}\sin\phi_{2}[(4\delta_{1}+\delta_{2})\omega^{2}\eta^{2}
+3\delta_{2}] -
6\eta^{2}\sin\theta_{1}\sin\phi_{1}\sin\theta_{3}\sin\phi_{3}
\nonumber\\
-\sqrt{2}\sin\theta_{1}\sin\phi_{1}(2\delta_{1}-\delta_{2})\omega\eta^{2}
- \sqrt{6}\sin\theta_{3}\sin\phi_{3}\delta_{2}\eta = 0 \,,
\end{eqnarray}
\begin{eqnarray}
i\sqrt{6}\delta_{2}\eta + i\sin\theta_{1}
\sin\phi_{1}6(1+\eta^{2})+\sin\theta_{2}
\cos\phi_{2}6(\eta^{2}-1)-\frac{\sqrt{6}}{2}\delta^{2}_{2}\eta
\nonumber\\
-\sin\theta_{1}\sin\phi_{1}[(2\delta_{1}+2\delta_{2})\eta^{2}+3\delta_{2}]
-6\omega^{2}\eta^{2}\sin\theta_{2}\sin\phi_{2}\sin\theta_{3}\sin\phi_{3}
\nonumber\\
-\sqrt{2}\sin\theta_{2}\sin\phi_{2}(2\delta_{1}-\delta_{2})\omega\eta^{2}
+\sqrt{3}\sin\theta_{3}\sin\phi_{3}\delta_{2}\omega\eta= 0 \,.
\end{eqnarray}
Firstly, the imaginary parts of the equations are in order of
$O(\delta_1,\delta_2)$, giving the same values of
$\sin\theta_{i}\sin\phi_{i}$ as in (\ref{sinsin2}). Contrarily, from
the real parts of the order of $O(\delta^{2}_1, \delta_1\delta_2,
\delta^2_2)$, we can obtain the expressions for
$\sin\theta_{i}\cos\phi_{i}$\,. These results read as
\begin{eqnarray}\label{high}
\sin\theta_{3}\cos\phi_{3} \simeq
-\frac{\sqrt{2}\omega}{36(1-\omega^{4})}\{2(1-\omega^{2})\delta^{2}_{1}
-2(1-\omega^{2})\delta_{1}\delta_{2}
\nonumber\\
- (1+2\omega^{2})\delta^{2}_{2} +
3\frac{(1+\omega^{2})(1+\eta^{2}+\omega^{2}\eta^{2})}
{(1+\eta^{2})(1+\omega^{2}\eta^{2})}\delta^{2}_{2}\}\,,
\\
\sin\theta_{2}\cos\phi_{2} \simeq
\frac{\sqrt{3}\omega\eta}{18(1-\omega^{4}\eta^{4})}\delta_{2}
(2\delta_{1}-\delta_{2})\{-\omega^{2}\eta^{2}
\nonumber\\
+ \frac{(1+\omega^{2}\eta^{2})(1+\eta^{2}+\omega^{2}\eta^{2})}
{(1+\omega^{2})(1+\eta^{2})}\}\,,
\\
\sin\theta_{1}\cos\phi_{1} \simeq -
\frac{\sqrt{6}\eta}{36(1-\eta^{4})}\delta_{2}
(2\delta_{1}-\delta_{2})\{-\eta^{2}
\nonumber\\
+ \omega^{2}\frac{(1+\eta^{2})(1+\eta^{2}+\omega^{2}\eta^{2})}
{(1+\omega^{2})(1+\omega^{2}\eta^{2})}\}\,.
\end{eqnarray}
Following the spirits of perturbation calculation, higher order
contributions only give small corrections, meaning that the results
are stable. Comparing these results with those of (\ref{sincos2}),
we can see that if the parameter $\omega$ or $\eta$(or both) is very
close to 1 (quasi-degenerate or inverted hierarchy mass spectrum),
the property of perturbation will be ruined. So a normal
hierarchy(small $\omega$ and $\eta$) neutrino mass spectrum is
favored in this scenario. Then for analytical expressions, we can
simplify them with the leading terms for power series of $\omega$
and $\eta$\,. For the results of $\sin\theta_{i}\sin\phi_{i}$ and
$\sin\theta_{i}\cos\phi_{i}$\,, we have
\begin{equation}
\sin\theta_{3}\sin\phi_{3} \simeq - \frac{\sqrt{2}}{6}\omega
(2\delta_{1}-\delta_{2})\;, \sin\theta_{2}\sin\phi_{2} \simeq
\frac{\sqrt{3}}{6}\omega\eta\delta_{2}\;, \sin\theta_{1}\sin\phi_{1}
\simeq -\frac{\sqrt{6}}{6}\eta\delta_{2}\;;
\end{equation}
and
\begin{eqnarray}
\sin\theta_{3}\cos\phi_{3} \simeq -\frac{\sqrt{2}}{18}\omega(
\delta^{2}_{1}-\delta_{1}\delta_{2}+ \delta^{2}_{2} )\,,
\nonumber\\
\sin\theta_{2}\cos\phi_{2} \simeq
\frac{\sqrt{3}}{18}\omega\eta\delta_{2} (2\delta_{1}-\delta_{2})\,,
\nonumber\\
\sin\theta_{1}\cos\phi_{1} \simeq -\frac{\sqrt{6}}{36}
\eta(\omega^{2}-\eta^{2}) \delta_{2}(2\delta_{1}-\delta_{2})\,.
\end{eqnarray}
Similarly, for the diagonal elements in
$\overline{M}^{}_{\nu}=V^{+}_{2}M^{(2)}_{\nu}V^{*}_{2}$\,, we can
get the information of $m_i$ and $\sigma_i$ as
\begin{eqnarray}
m_{1}\simeq \frac{z^{2}\omega^{2}\eta^{2}}{M^{}_{0}}(1 + \frac{1}{9}
\frac{1-\omega^{2}}{1+\omega^{2}}\delta^{2}_{1} -
\frac{1}{9}\frac{1-\omega^{2}}{1+\omega^{2}}\delta_{1}\delta_{2} +
\frac{1}{72}\frac{5+\omega^{2}-\omega^{2}\eta^{2}(1+5\omega^{2})}
{(1+\omega^{2})(1+\omega^{2}\eta^{2})}\delta^{2}_{2})\,,
\nonumber\\
m_{2} \simeq \frac{z^{2}\eta^{2}}{M^{}_{0}}(1 -
\frac{1}{9}\frac{1+3\omega^{2}}{1+\omega^{2}} \delta^{2}_{1} +
\frac{1}{9}\frac{1+3\omega^{2}}{1+\omega^{2}}\delta_{1}\delta_{2} +
\frac{1}{18}\frac{1+4\eta^{2}+3\omega^{2}\eta^{2}}
{(1+\omega^{2})(1+\eta^{2})}\delta^{2}_{2})\,,
\nonumber\\
m_{3}\simeq \frac{z^{2}}{M^{}_{0}}(1 - \frac{1}{8}\delta^{2}_{2} +
\frac{1}{12}\eta^{2}\frac{2+\omega^{2}+3\omega^{2}\eta^{2}}
{(1+\eta^{2})(1+\omega^{2}\eta^{2})}\delta^{2}_{2})\,.
\end{eqnarray}
and
\begin{equation}
\tan2\sigma_{1} \simeq \frac{1}{6}(4\delta_{1}+\delta_{2})\;,\;
\tan2\sigma_{2} \simeq \frac{1}{3}(\delta_{1}+\delta_{2})\;,\;
\tan2\sigma_{3} \simeq \frac{1}{2}\delta_{2}\;.
\end{equation}
For small mass ratios, the three masses are expressed as
\begin{eqnarray}
m_{1} \simeq \frac{z^{2}\omega^{2}\eta^{2}}{M^{}_{0}}(1 +
\frac{1}{9} \delta^{2}_{1} - \frac{1}{9}\delta_{1}\delta_{2} +
\frac{5}{72} \delta^{2}_{2})\,,
\nonumber\\
m_{2} \simeq \frac{z^{2}\eta^{2}}{M^{}_{0}}(1 - \frac{1}{9}
\delta^{2}_{1} + \frac{1}{9}\delta_{1}\delta_{2} + \frac{1}{18}
\delta^{2}_{2})\,,
\nonumber\\
m_{3} \simeq \frac{z^{2}}{M^{}_{0}}(1 - \frac{1}{8}
\delta^{2}_{2})\;.
\end{eqnarray}

\newpage
\begin{figure}
\begin{center}
\vspace{0cm}
\epsfig{file=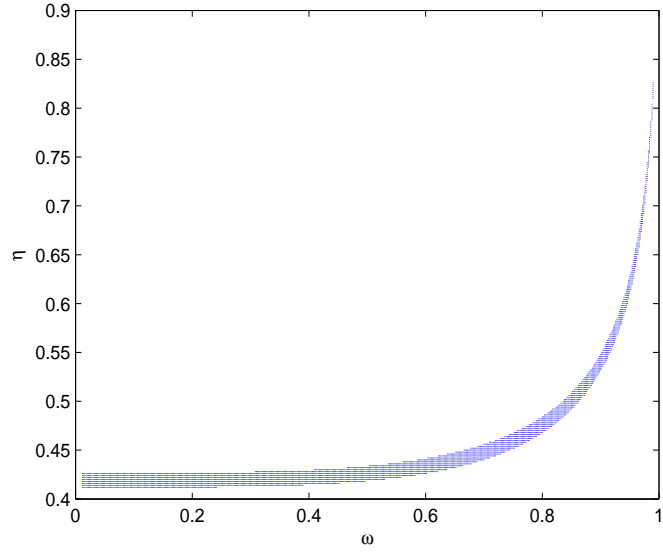, 
width=10cm, height=8cm, angle=0, clip=0} \caption{Allowed parameter
space of $\omega$ and $\eta$ from $1\sigma$ region of the two
$\Delta m^2$.}
\end{center}
\end{figure}

\begin{figure}
\begin{center}
\vspace{0cm}
\epsfig{file=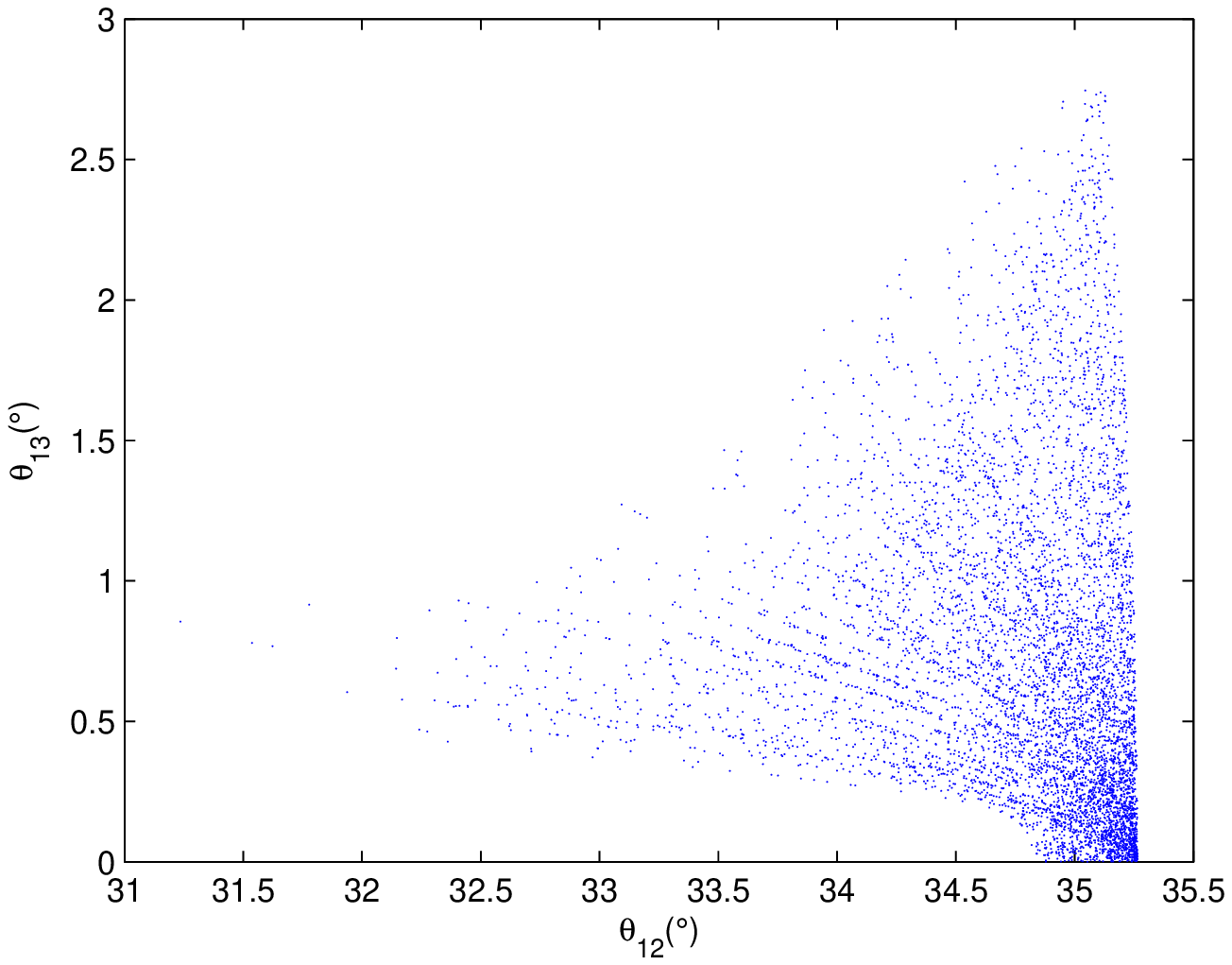, 
width=10cm, height=8cm, angle=0, clip=0} \caption{Allowed parameter
distribution of $\theta_{12}$ and $\theta_{13}$ for ($\delta_{1}$
$\delta_{2}$) within \,[\,--\,0.5\,,\,0.5\,].}
\end{center}
\end{figure}

\newpage

\begin{figure}
\begin{center}
\vspace{0cm}
\epsfig{file=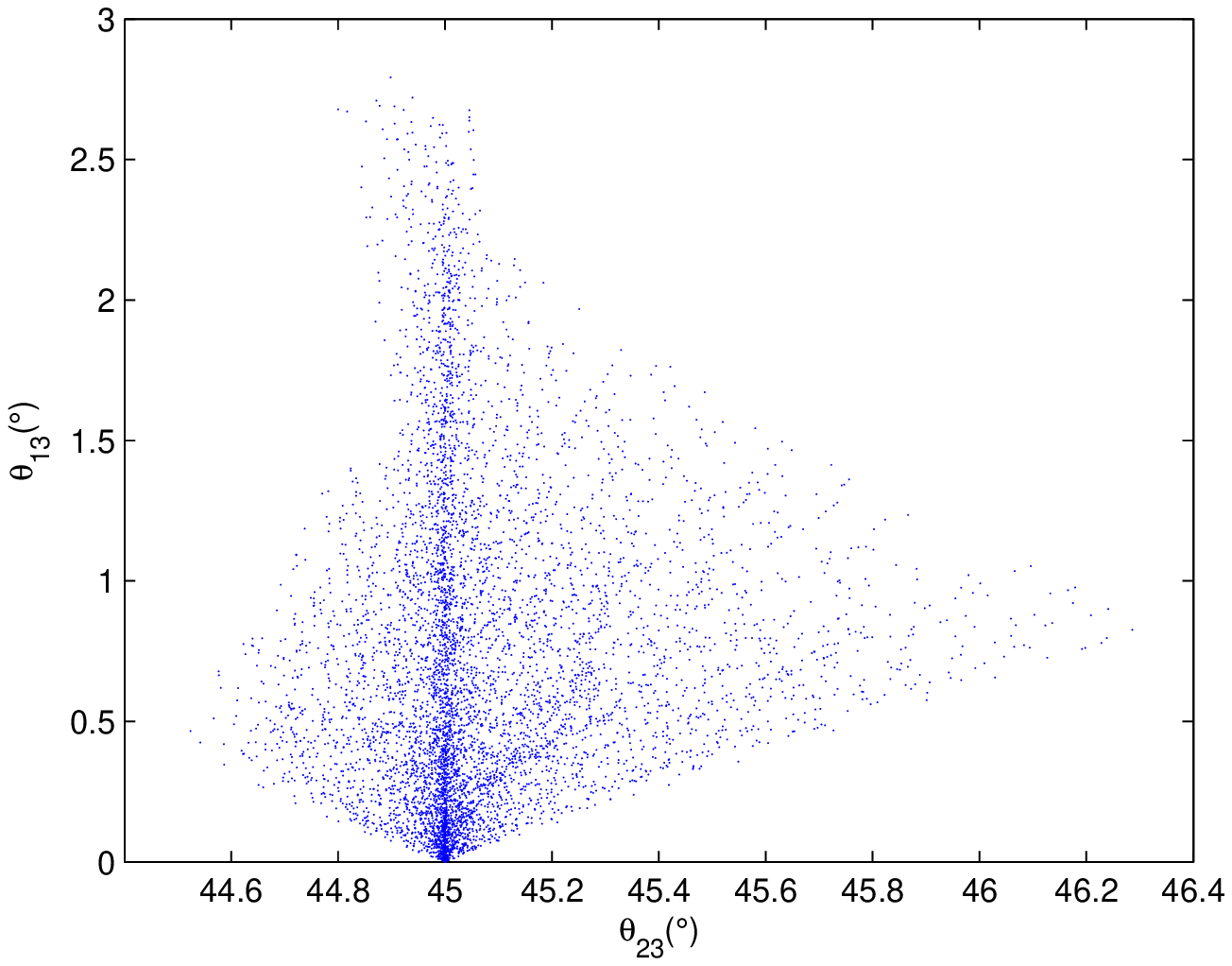, 
width=10cm, height=8cm, angle=0, clip=0} \caption{Allowed parameter
distribution of $\theta_{23}$ and $\theta_{13}$ for ($\delta_{1}$
$\delta_{2}$) within \,[\,--\,0.5\,,\,0.5\,]}
\end{center}
\end{figure}

\begin{figure}
\begin{center}
\vspace{0cm}
\epsfig{file=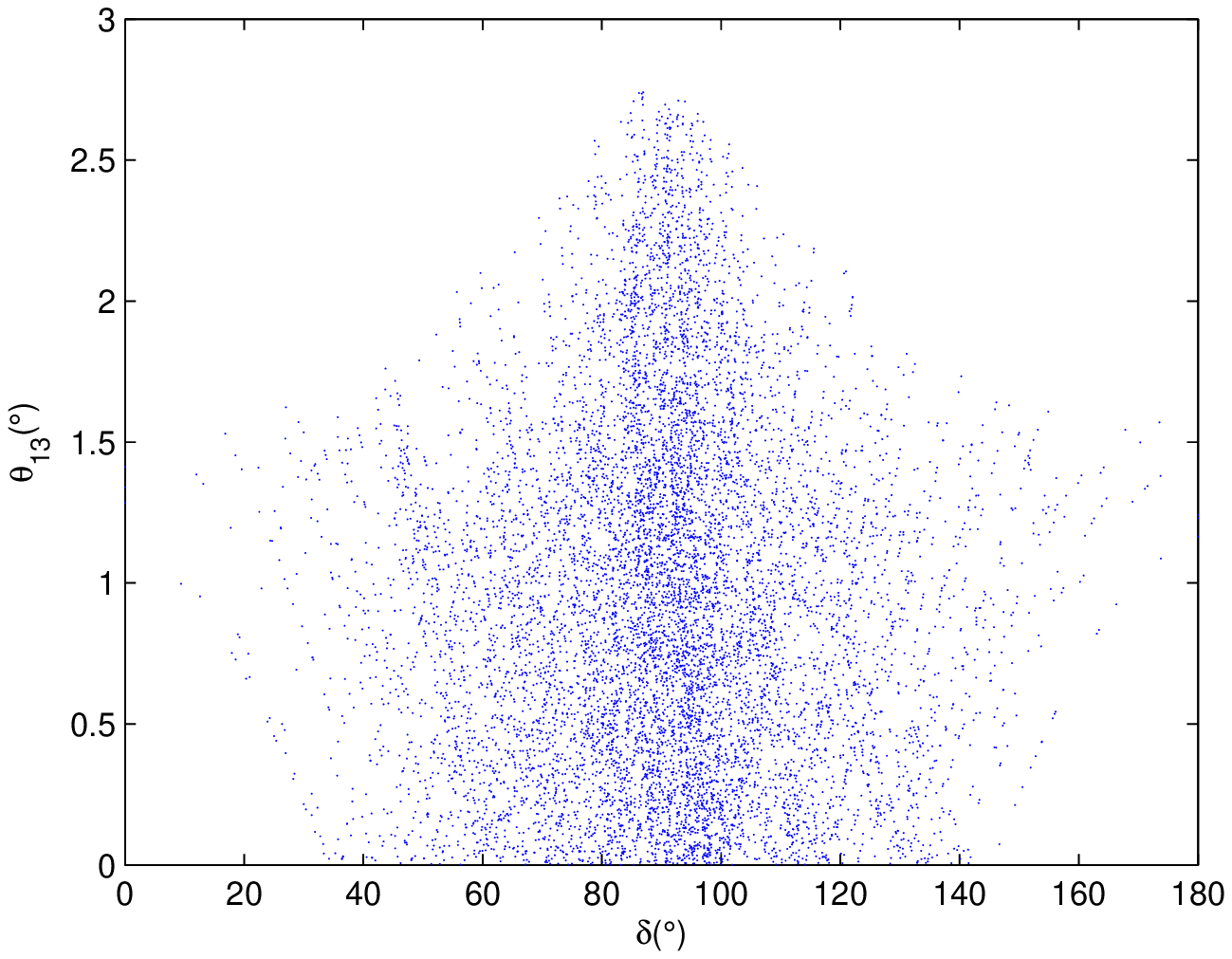, 
width=10cm, height=8cm, angle=0, clip=0} \caption{Allowed parameter
distribution of $\delta$ and $\theta_{13}$ for ($\delta_{1}$
$\delta_{2}$) within \,[\,--\,0.5\,,\,0.5\,] .}
\end{center}
\end{figure}

\newpage

\begin{figure}
\begin{center}
\vspace{0cm}
\epsfig{file=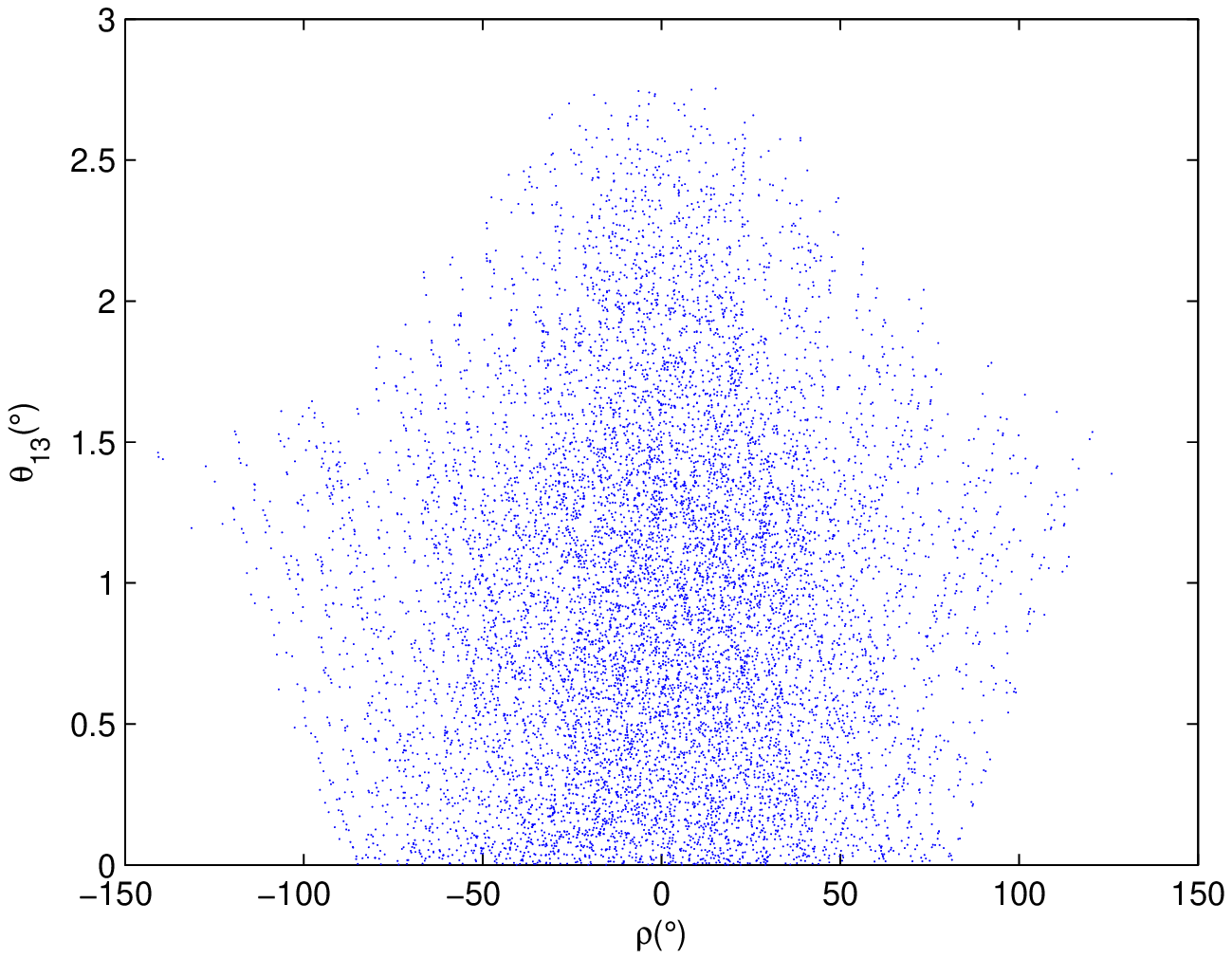, 
width=10cm, height=8cm, angle=0, clip=0} \caption{Allowed parameter
distribution of $\rho$ and $\theta_{13}$ for ($\delta_{1}$
$\delta_{2}$) within \,[\,--\,0.5\,,\,0.5\,] .}
\end{center}
\end{figure}

\begin{figure}
\begin{center}
\vspace{0cm}
\epsfig{file=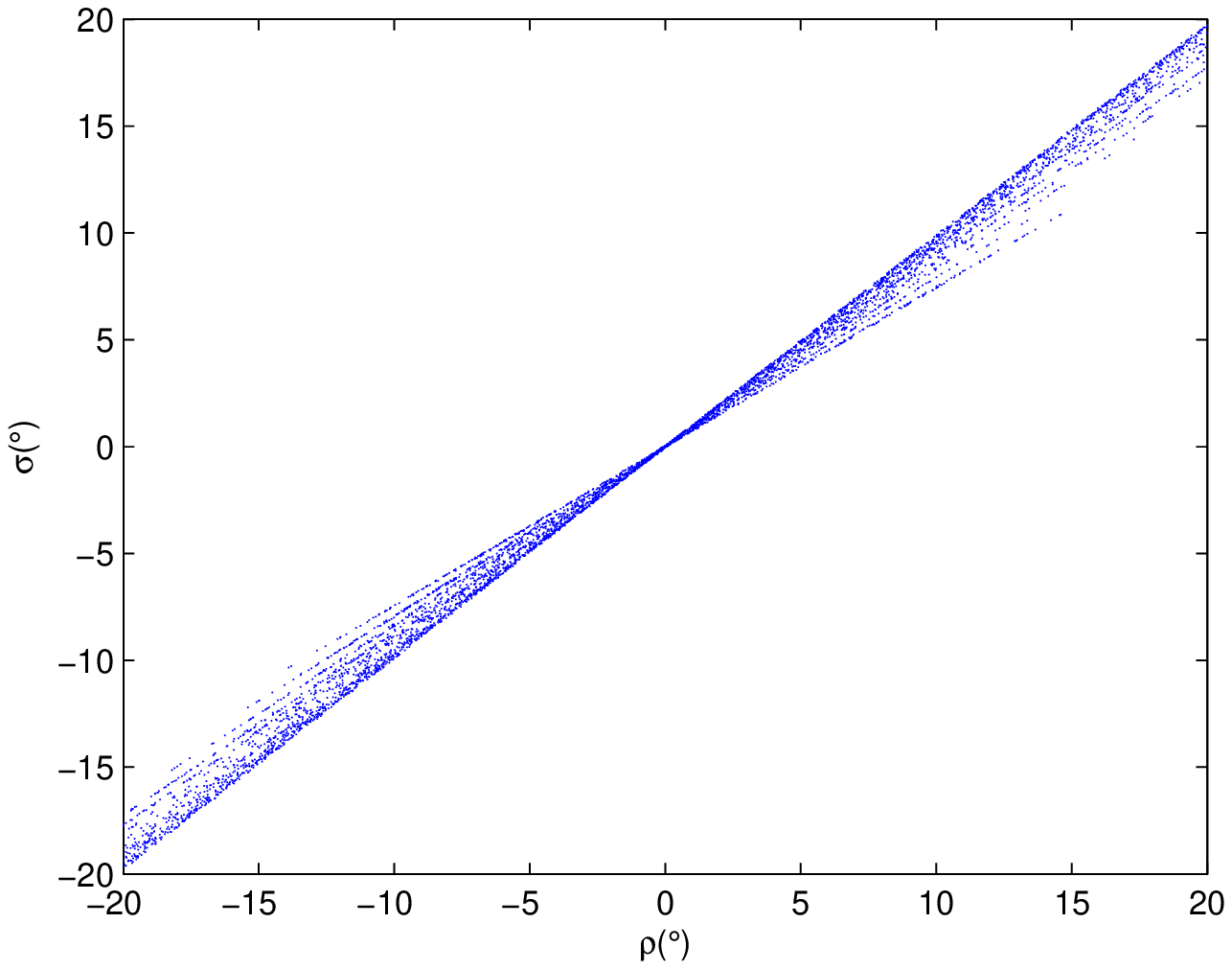, 
width=10cm, height=8cm, angle=0, clip=0} \caption{Parameter
dependence of $\rho$ and $\sigma$ for ($\delta_{1}$ $\delta_{2}$)
within \,[\,--\,0.5\,,\,0.5\,] .}
\end{center}
\end{figure}


\begin{thebibliography}{99}

\bibitem{sno1}
Q.R.~Ahmad {\it et al.} [SNO Collaboration], Phys.\ Rev.\ Lett.\
{\bf 87}, (2001) 071301.
\bibitem{sno2}
Q.R.~Ahmad {\it et al.} [SNO Collaboration], Phys.\ Rev.\ Lett.\
{\bf 89}, (2002) 011301.
\bibitem{sno3}
Q.R.~Ahmad {\it et al.} [SNO Collaboration], Phys.\ Rev.\ Lett.\
{\bf 89}, (2002) 011302.
\bibitem{sno4}
Q.R.~Ahmad {\it et al.} [SNO Collaboration], Phys.\ Rev.\ Lett.\
{\bf 92}, (2004) 181301.
\bibitem{sno5}
Q.R.~Ahmad {\it et al.} [SNO Collaboration], Phys.\ Rev.\ Lett.\
{\bf 101}, (2008) 111301.

\bibitem{Kamiokande1}
K.S.~Hirata, {\it et al.}, Phys.\ Rev.\ D {\bf 44} (1991) 2241.
\bibitem{Kamiokande2}
K.S.~Hirata, {\it et al.}, Phys.\ Rev.\ Lett.\ {\bf 66}, (1991) 9.
\bibitem{Kamiokande3}
Y.~Fukuda, {\it et al.}, Phys.\ Rev.\ Lett.\ {\bf 77}, (1996) 1683.


\bibitem{SK1}
Y.~Fukuda {\it et al.}, [Super-Kamiokande Collaboration], Phys.\
Rev.\ Lett.\ {\bf 87} (1998) 1562.
\bibitem{SK2}
Y.~Ashie {\it et al.}, [Super-Kamiokande Collaboration], Phys.\
Rev.\ Lett.\ {\bf 93} (2004) 101801, [hep-ex/0404034].
\bibitem{SK3}
Y.~Ashie {\it et al.} [Super-Kamiokande Collaboration], Phys.\ Rev.\
D {\bf 71},(2005) 112005, [hep-ex/0501064].

\bibitem{kldet1}
K.~Eguchi {\it et al.} [KamLAND Collaboration], Phys.\ Rev.\ Lett.\
{\bf 90}, (2003) 021802.
\bibitem{kldet2}
T.~Araki {\it et al.} [KamLAND Collaboration], Phys.\ Rev.\ Lett.\
{\bf 94}, (2005) 081801.
\bibitem{kldet3}
T.~Araki {\it et al.} [KamLAND Collaboration], Phys.\ Rev.\ Lett.\
{\bf 100}, (2008) 221803.

\bibitem{CHOOZ}
M.~Apollonio, {\it et al.} [CHOOZ Collaboration], Eur.\ Phys.\ J.\ C
{\bf 27}, (2003) 331.

\bibitem{K2K1}
M.H.~Alm {\it et al.} [K2K Collaboration], Phys.\ Rev.\ Lett.\ {\bf
90} (2003) 041801.
\bibitem{K2K2}
M.H.~Ahn {\it et al.} [K2K Collaboration], Phys.\ Rev.\ D {\bf 74}
(2006) 072003.

\bibitem{MINOS1}
P.~Adamson {\it et al.} [MINOS Collaboration] Phys.\ Rev.\ Lett.\
{\bf 101} (2008) 131802.


\bibitem{PDG}
C.~Amsler {\it et al.} [Particle Data Group], Phys. Lett. B {\bf
667}, 1 (2008).

\bibitem{SV}
A.~Strumia and F.~Vissani, arXiv:hep-ph/0606054v2.

\bibitem{GF1}
T.~Schwetz, M.~Tortola and J.W.F.~Valle, New J. Phys. {\bf 10}
(2008) 113011 [arXiv:0808.2016].
\bibitem{GF2}M.~Maltoni and T.~Schwetz, arXiv:0812.3161 [hep-ph].

\bibitem{TB1}
P.F.~Harrison, D.H.~Perkins and W.G.~Scott, Phys.\ Lett.\ B {\bf
530}, 167 (2002).
\bibitem{TB2}
P.F.~Harrison and W.G.~Scott, Phys.\ Lett.\ B {\bf 535}, 163 (2002).
\bibitem{TB3}
Z.Z.~Xing, Phys.\ Lett.\ B {\bf 533}, 85 (2002).
\bibitem{TB4}
X.G.~He and A.~Zee,
Phys.\ Lett.\ B {\bf 560}, 87 (2003).

\bibitem{FS1}
R.N.~Mohapatra and A.Y.~Smirnov, Ann.\ Rev.\ Nucl.\ Part.\ Sci.\
{\bf 56}, 569 (2006), [arXiv:hep-ph/0603118].
\bibitem{FS2}
E.~Ma, J.\ Phys.\ Conf.\ Ser.\ {\bf 53}, 451 (2006),
[arXiv:hep-ph/0606024].

\bibitem{TB51}
S.F.~King, J.\ High\ Energy\ Phys.\ {\bf 05} 08 (2005),
[arXiv:hep-ph/0506297].
\bibitem{TB52}
F.~Plentinger and W.~Rodejohann, Phys.\ Lett.\ B {\bf 625} 264
(2005), [arXiv:hep-ph/0507143].
\bibitem{TB53}
K.A.~Hochmuth, S.T.~Petcov and W.~Rodejohann, Phys.\ Lett.\ B {\bf
654} 177 (2007), arXiv:0706.2975 [hep-ph].

\bibitem{TB61}
S.~Luo and Z.Z.~Xing, Phys.\ Lett.\ B {\bf 632} 341 (2006),
[arXiv:hep-ph/0509065].
\bibitem{TB62}
M.~Hirsch, E.~Ma, J.C.~Romao, J.W.F.~Valle and A.V.del~Moral Phys.\
Rev.\  D {\bf 75} 053006 (2007), [arXiv:hep-ph/0606082].
\bibitem{TB63}
A.~Dighe, S.~Goswami and W.~Rodejohann, Phys.\ Rev.\ D {\bf 75}
073023 (2007), [arXiv:hep-ph/0612328].


\bibitem{SS}
P.~Minkowski, Phys.\ Lett.\ B {\bf 67}, 421 (1977).
\bibitem{SS1}
T.~Yanagida, in {\it Proceedings of the Workshop on Unified Theory
and the Baryon Number of the Universe}, edited by O.~Sawada and
A.~Sugamoto (KEK, Tsukuba, 1979), p.~95.
\bibitem{SS2}
M.~Gell-Mann, P.~Ramond, and R.~Slansky, in {\it Supergravity},
edited by F.~van Nieuwenhuizen and D.~Freedman (North Holland,
Amsterdam, 1979), p.~315.
\bibitem{SS3}
S.L.~Glashow, in {\it Quarks and Leptons}, edited by
M.~L$\rm\acute{e}vy$ {\it et al.} (Plenum, New York, 1980), p.~707.
\bibitem{SS4}
R.N.~Mohapatra and G.~Senjanovic, Phys.\ Rev.\ Lett.\ {\bf 44}, 912
(1980).


\bibitem{GL1}
W.~Grimus and L.~Lavoura, J.\ High\ Energy\ Phys.\ {\bf 07}, 045
(2001).
\bibitem{GL2}
W.~Grimus and L.~Lavoura, Phys.\ Lett.\ B {\bf 572}, 189 (2003).
\bibitem{GL3}
W.~Grimus and L.~Lavoura, J.\ High\ Energy\ Phys.\ {\bf 08}, 013
(2005).


\bibitem{KX1} S.K.~Kang and C.S.~Kim, Phys.\ Lett.\ B {\bf 634},(2006)
520, arXiv:hep-ph/0511106.
\bibitem{KX2}
S.K.~Kang Z.Z.~Xing and S.~Zhou, Phys.\ Rev.\ D {\bf 73} (2006)
013001, arXiv:hep-ph/0511157.

\bibitem{J}
C.~Jarlskog, Phys.\ Rev.\ Lett.\ {\bf 55}, 1039 (1985).
\bibitem{Wu}
D.D.~Wu, Phys.\ Rev.\ D {\bf 33}, 860 (1986).

\bibitem{deg1}
G.~C.~Branco, M.~N.~Rebelo, and J.~I.~Silva-Marcos, Phys.\ Rev.\
Lett.\ {\bf 82}, 683 (1999).
\bibitem{deg2}
R.~Adhikari, E.~Ma, and G.~Rajasekaran, Phys.\ Lett.\ B {\bf 486},
134 (2000).

\bibitem{LFV} G.C.~Branco, M.N.~Rebelo and J.I.~Silva-Marcos,
Phys.\ Lett.\ B {\bf 633} 345 (2006), arXiv:hep-ph/0510412v2.

\bibitem{FY} M.~Fukugita and T.~Yanagida, Phys.\ Lett.\ B {\bf 174} 45 (1986).

\bibitem{HE} X.G.~He and A.~Zee, Phys.\ Lett.\ B {\bf 645}, 427
(2007), arXiv:hep-ph/0607163v3.
\bibitem{Tanimoto}
M.~Honda and M.~Tanimoto, Prog.\ Theor.\ Phys.\ {\bf 119} 583
(2008), arXiv:0801.0181v2 [hep-ph].

\bibitem{PA1}
N.~Li and B.Q.~Ma, Phys.\ Rev.\ D {\bf 71} 017302 (2005),
arXiv:hep-ph/0412126v2.
\bibitem{PA2} S.F.~King, Phys.\ Lett.\ B {\bf 659}, 244 (2008),
arXiv:0710.0530v3 [hep-ph].
\bibitem{PA3}
S.~Pakvasa, W.~Rodejohann and T.~Weiler, Phys. Rev. Lett. {\bf 100},
111801 (2008), arXiv:0711.0052v2 [hep-ph].

\bibitem{DCH} Double Chooz Collaboration, arXiv:hep-ex/0606025.
\bibitem{daya} Daya Bay Collaboration, arXiv:hep-ex/0701029.


\end{thebibliography}
\end{document}